\documentclass[useAMS,usenatbib]{mn2e}
\usepackage{epsfig}
\usepackage{color} 
\usepackage{natbib}
\def\mgii{Mg~{\sc ii}~}

\def\civ{C~{\sc iv}~}
\def\siiv{Si~{\sc iv}~}

\def\kms{km~s$^{-1}$} 
\title[Microvariability properties of radio-loud BALQSOs]{Intranight optical variability of radio-loud broad
  absorption line quasars } \author[Joshi \& Chand]{Ravi Joshi$^{1}$\thanks{E-mail: ravi@aries.res.in (RJ);
    hum@aries.res.in (HC)}, Hum Chand$^{1}$  \\ $^{1}$Aryabhatta Research Institute of
  Observational Sciences (ARIES), Manora Peak, Nainital, 263129,
  India}

\begin{document}
\date{Accepted ---. Received ---; in original form ---}

\pagerange{\pageref{firstpage}--\pageref{lastpage}} \pubyear{2012}

\maketitle

\label{firstpage}
\begin{abstract}
We present the results of an optical photometric monitoring program of
 10 extremely radio loud broad absorption line quasars
(RL-BALQSOs) with radio-loudness parameter, $R$, greater than 100 and
magnitude g$_{i} < 19$. Over an observing run of about 3.5-6.5 hour 
we found a clear detection of variability for one of
our 10  radio-loud BALQSOs with the INOV duty cycle of  5.1 per cent,
while on including the probable variable cases, a higher duty cycle of
35.1 per cent is found; which are very similar to the duty cycle of
radio quiet broad absorption line quasars (RQ-BALQSOs). This low duty
cycle of clear variability per cent in radio-loud sub-class of BALQSOs
can be understood under the premise where BALs outflow may arise from
large variety of viewing angles from the jet axis or perhaps being closer
to the disc plane.

\end{abstract}


%%%%%%%%%%%%%%%%%%%%%%%%%%%%%%%%%
%%%%%%%%% Introduction  %%%%%%%%%
%%%%%%%%%%%%%%%%%%%%%%%%%%%%%%%%%

\section{Introduction}
\label{sec:intro_rlbal}

Even after four decades of intensive research the physical nature of
quasar variability is an open question. The quasar optical variability
has long been a tool used to set limits on the size of the emitting
regions, to constrain emission models, and to probe physical
conditions close to the central black hole in active galactic nuclei
(AGN). The microvariability phenomena found in radio-loud sub classes
of AGN with $R$\footnote{Radio-loudness parameter, $R$, is defined as
  the ratio of radio [5 GHz] flux to the optical [2500~\AA]
  flux.}$>10$, is quite common and is widely believed to be connected
to the conditions in relativistic jets. Since radio-quiet quasars (RQQSOs) lack jets of
significant power and extent, the microvariability seen in them may
arise from processes on the accretion disc itself, and thus could be
used to probe the discs
~\citep*[e.g.][]{GopalKrishna1993MNRAS.262..963G}. Our current
understanding of optical microvariability is that only about $\sim$10
per cent of the radio-quiet quasars, which constitute $\sim$85$-$90
per cent of the quasar population, show microvariation, while rest of
10 per cent radio-loud (RL) sub-class shows about 35$-$40 per cent
microvariability, when observed continuously for $\sim$6 hour
~\citep{GopalKrishna2003ApJ...586L..25G,
  Stalin2004MNRAS.350..175S,Gupta2005A&A...440..855G}. Recently, the
microvariability studies have been extended to understand the nature
of the radio-quiet quasar sub-class with broad absorption lines
(BALs), the RQ-BALQSOs ~\citep{Joshi2011MNRAS.412.2717J}, but still
there has been a lack of systematic efforts towards the
microvariability properties of their radio-loud counter part, i.e.
radio-loud broad absorption lines, the RL-BALQSOs. \par

The BAL quasars are characterized by the presence of strong absorption
troughs, attributed to material flowing outwards from the nucleus with
velocities of 5000 to 50000 \kms 
~\citep{Green2001ApJ...558..109G}. They constitute about 10$-$15 per
cent of optically selected quasars
\citep*[e.g.][]{Reichard2003AJ....126.2594R,Hewett2003AJ....125.1784H};
which is generally interpreted as to represent the covering factor of
an outflowing BAL wind. In addition, BALQSOs are primarily believed to
belong the class of radio-quiet QSOs
~\citep{Stocke1992ApJ...396..487S}. However with the advent of large
comprehensive radio surveys, it has become clear that BALQSOs also
constitute a significant fraction of luminous radio quasar
population ~\citep{Becker2000ApJ...538...72B}. \par

In term of microvariation of BALQSOs, recent compilation of
microvariability studies by~\citet{Carini2007AJ....133..303C} has
pointed out that the BALQSOs may be an interesting class, as these
shows about 50 per cent microvariation similar to the blazars,
although their conclusion was based on just 6 RQ-BALQSOs. This was further
investigated by ~\citet{Joshi2011MNRAS.412.2717J} with three times
larger sample size, and found that the true fraction of RQ-BALQSOs
microvariation is just 2 out of 19, which was found similar to the
usual low microvariability fraction of normal RQQSOs with observation
lengths of about 4 hour. This result has rather provided support for
models where radio-quiet BALQSOs do not appear to be a special case of
the RQQSOs in terms of their microvariability properties. Further,
it has also given support to the model, where BAL outflow appear to be
closer to the disk plane ~\citep[e.g.][]{Elvis2000ApJ...545...63E}.
\par

 In addition to probe the difference of microvariability properties
 among various class of AGN, there has been also considerable
 systematic efforts to investigate whether it depends on the
 radio-loudness of the AGN or not. For instance,
 ~\citet{Diego1998ApJ...501...69D} have performed photometric optical
 observations of a sample of 17 core-dominated radio-loud quasars
 (CRL-QSOs) and 17 radio-quiet quasars (RQQSOs), to compare their
 microvariability properties. They found that microvariation in
 RQQSOs may be as frequent as in CRL-QSOs, which was also augmented by
 a similar result of ~\citet{Ramirez2009AJ....138..991R}. However,
 still there have been a lack of such systematic efforts to know whether
 or not this similarity also extend among the RQ-BALQSOs and RL-BALQSOs. 
 This form the main motivation of our present work; to
 characterize for the first time microvariability properties of
 radio-loud BALQSOs and compare them with our earlier study of the
 radio-quiet BALQSOs ~\citep{Joshi2011MNRAS.412.2717J}, using the same
 analysis and statistical method in both the cases. The investigation
 would also have an important implication in understanding the role of
 radio-loudness in microvariability properties and to get a clue whether
 the cause of microvariation in radio-loud and radio-quiet BAL quasars
 are similar or not. \par

Paper is organized as follows: in Section 2, we describe our
sample selection criteria, our observations and the data reductions
while in, Section 3, we describe our analysis and results. Finally,
discussion and conclusion are given in Section 4.

%%%%%%%%%%%%%%%%%%%%%%%%%%%%%%%%%%%%%
%%%%%%%%% Sample selection  %%%%%%%%%
%%%%%%%%%%%%%%%%%%%%%%%%%%%%%%%%%%%%%

\section{Sample selection}
\label{sec:sample_selection_rlbal}

We select a sample of 10 radio-loud BALQSO from the compilation of
~\citet{Shen2011ApJS..194...45S} (their online Table 1) based on Sloan
Digital Sky survey, Data Release 7 (SDSS DR$-$7;
~\citet{Abazajian2009ApJS..182..543A}) quasar catalog
~\citep{Schneider2010yCat.7260....0S}, which satisfy two of our main
selection criteria. First, the radio-loudness parameter, $R$, 
should be greater than 100. Here, the criterian chosen to limit
radio-loudness parameter, R $>$ 100, was based on our requirement
that, sample size should be large enough for the sake of better
statistics and the selected sources should have reasonably high
radio-loudness, so as to help in detection of any effect being mainly
due to the difference in radio-loudness, between the two sub-classes
i.e. the radio-loud BALQSOs verses the earlier studied radio-quiet
BALQSOs ~\citep{Joshi2011MNRAS.412.2717J}. Second, the source need to
be bright enough with mag g$_{i} < 19$, so that even with a 1-2 m class
telescope we could obtain a good enough signal-to-noise ratio (SNR) to
detect fluctuations of $<0.02$ mag with a reasonably good time
resolution of $< 10$ minute. \par

We also limit the RL-BALQSOs to have absolute magnitudes 
M$_{i}<-24.5$, so that the flux contribution from the host galaxy can be 
assumed to be negligible ~\citep{Miller1990MNRAS.244..207M}. Our final
sample consists of a total of 10 RL-BALQSOs, with  3 lobe and 7 core
dominated radio morphology ~\citep[][]{Schneider2010AJ....139.2360S, Shen2011ApJS..194...45S}, as listed in Table~\ref{tab:sample}. The
whole sample covers a redshift range of  $ 0.52 \le z_{em} \le 3.06$.

\begin{table*}
\centering
\begin{minipage}{180mm}
\caption{Properties of the observed radio-loud BALQSOs.}
\label{tab:sample}
\begin{tabular}{@{}lll lll ll @{}}
\multicolumn{1}{c}{Object}  &\multicolumn{1}{c}{$\alpha_{2000.0}$}
&\multicolumn{1}{c}{$\delta_{2000.0}$} &{g$_{i}$} &{M$_{i}$}
&{$z_{em}$} &{R\footnote{Ratio of the radio [5 GHz] flux to the
    optical [2500\AA] flux taken from SDSS DR7 ~\citep{Schneider2010yCat.7260....0S}.}}  &{Type\footnote{Quasar
    type: 1=core-dominant; 2=lobe-dominant (radio morphology
    classification following ~\citealt{Jiang2007ApJ...656..680J}).      }} \\
\hline 
 J004323.42$-$001552.4  &   00$^h$ 43$^m$ 23.42$^s$& $-$00$^{\circ}$ 15$^{'}$ 52.48$^{''}$ & 18.80 &$-27.0$& 2.797& $597.97$ & 2\\
 J082231.53$+$231152.0  &   08$^h$ 22$^m$ 31.53$^s$& $+$23$^{\circ}$ 11$^{'}$ 52.00$^{''}$ & 17.94 &$-25.2$& 0.653& $117.12$ & 2\\
 J085641.56$+$424253.9  &   08$^h$ 56$^m$ 41.56$^s$& $+$42$^{\circ}$ 42$^{'}$ 53.90$^{''}$ & 18.83 &$-27.9$& 3.061& $198.53$ & 1\\
 J092913.96$+$375742.9  &   09$^h$ 29$^m$ 13.96$^s$& $+$37$^{\circ}$ 57$^{'}$ 42.90$^{''}$ & 18.32 &$-27.8$& 1.915& $147.51$ & 1\\
 J095327.95$+$322551.6  &   09$^h$ 53$^m$ 27.95$^s$& $+$32$^{\circ}$ 25$^{'}$ 51.00$^{''}$ & 17.77 &$-27.7$& 1.574& $278.62$ & 2\\
 J112938.46$+$440325.0  &   11$^h$ 29$^m$ 38.46$^s$& $+$44$^{\circ}$ 03$^{'}$ 25.00$^{''}$ & 18.25 &$-27.8$& 2.211& $250.81$ & 1\\
 J115944.82$+$011206.9  &   11$^h$ 59$^m$ 44.82$^s$& $+$01$^{\circ}$ 12$^{'}$ 06.00$^{''}$ & 17.58 &$-28.4$& 2.002& $471.59$ & 1\\
 J121539.66$+$090607.4  &   12$^h$ 15$^m$ 39.66$^s$& $+$09$^{\circ}$ 06$^{'}$ 07.40$^{''}$ & 18.37 &$-27.9$& 2.722& $156.77$ & 1\\
 J122848.21$-$010414.4  &   12$^h$ 28$^m$ 48.21$^s$& $-$01$^{\circ}$ 04$^{'}$ 14.40$^{''}$ & 18.31 &$-28.1$& 2.655& $119.09$ & 1\\
 J160354.14$+$300208.6  &   16$^h$ 03$^m$ 54.14$^s$& $+$30$^{\circ}$ 02$^{'}$ 08.60$^{''}$ & 18.16 &$-27.6$& 2.031& $185.66$ & 1\\
\hline                                                            
\end{tabular}                                                    
\end{minipage}
\end{table*}

%%%%%%%%%%%%%%%%%%%%%%%%%%%%%%%%%%%%%
%%%%%%  Photometric observations %%%%
%%%%%%%%%%%%%%%%%%%%%%%%%%%%%%%%%%%%%

 \subsection{Photometric observations}
\label{sec:photometry_rlbal}
The photometric monitoring is carried out by using the 1.3-m Devasthal
fast optical telescope (hereafter 1.3-m DFOT) operated by Aryabhatta Research Institute of
observational sciencES (ARIES), Nainital, India. DFOT is a fast beam
(f/4) optical telescope with pointing accuracy better than 10 arcsec
RMS ~\citep{Sagar2011Csi...101...8.25}. The telescope is equipped with
Andor CCD having 2048 $\times$ 2048 pixels, with pixel size of 13.5
micron resulting in the field of view of 18 arcmin on the sky. The CCD is
read out with 31 and 1000 kHz speed, with the respective system RMS
noise of 2.5, 7 e$^{-}$ and gain of 0.7, 2 e${^-}$/ADU (Analog to
Digital Unit). The camera is cooled down thermoelectrically to $-$85
$^{\circ}$C. We perform continuous monitoring of each source for about
3.5$-$6.5 hour in SDSS$-$r passband at which our CCD system has maximum
sensitivity. We observe each science frame for about 5$-$10 minute, to
achieve typical SNR grater than 25. The typical seeing during our
observing runs was 1.5$-$3.8 arcsec.

\subsection{Data Reduction}

 All image pre-processing steps, i.e. bias subtraction, flat-fielding
 and cosmic-ray removal are carried out using the standard tasks
 available in the data reduction software {\textsc
   IRAF} \footnote{\textsc {Image Reduction and Analysis Facility
     (http://iraf.noao.edu/) }}. The instrumental magnitudes of the
 comparison stars and the target source are obtained from the data by
 using Dominion Astronomical Observatory Photometry\textrm{II}
 (DAOPHOT II)
 \footnote{\textsc {Dominion Astrophysical Observatory Photometry}}
 software to perform the concentric circular aperture photometric
 technique ~\citep{Stetson1987PASP...99..191S}. Aperture photometry
 is carried out with four aperture radii, 1, 2, 3, \& 4 times of the full
 width at half maximum (FWHM). Utmost caution has been taken to deal
 with the seeing and for that, we have taken the mean FWHM of 5 fairly
 bright stars on each CCD frame in order to choose the aperture for
 the photometry of that frame. The data reduced with
 different aperture radii are found to be in good agreement. However,
 we adopted the aperture radii of $\sim$2$\times$FWHM for our final
 results as it has almost always provided the best signal-to-noise
 ratio. \par

To derive the Differential Light Curves (DLCs) of quasar, we have
selected two steady comparison stars from the same field, on the basis
of their proximity in both location and magnitude to the quasar. The
locations of the two best comparison stars for each RL-BALQSO is
given in columns 3, 4 of Table ~\ref{tab_cdq_comp}. The typical colour
differences $g-r$ for our quasar-star and star-star pairs are close to
unity (see column 7, Table ~\ref{tab_cdq_comp}). A detailed
investigation quantifying the effect of colour differences by
~\citet{Carini1992AJ....104...15C} \&
~\citet{Stalin2004MNRAS.350..175S} shows that the effect of colour
differences of this amount on DLCs will be negligible for a specific
band. \par

\begin{figure*} 
\centering
\epsfig{figure=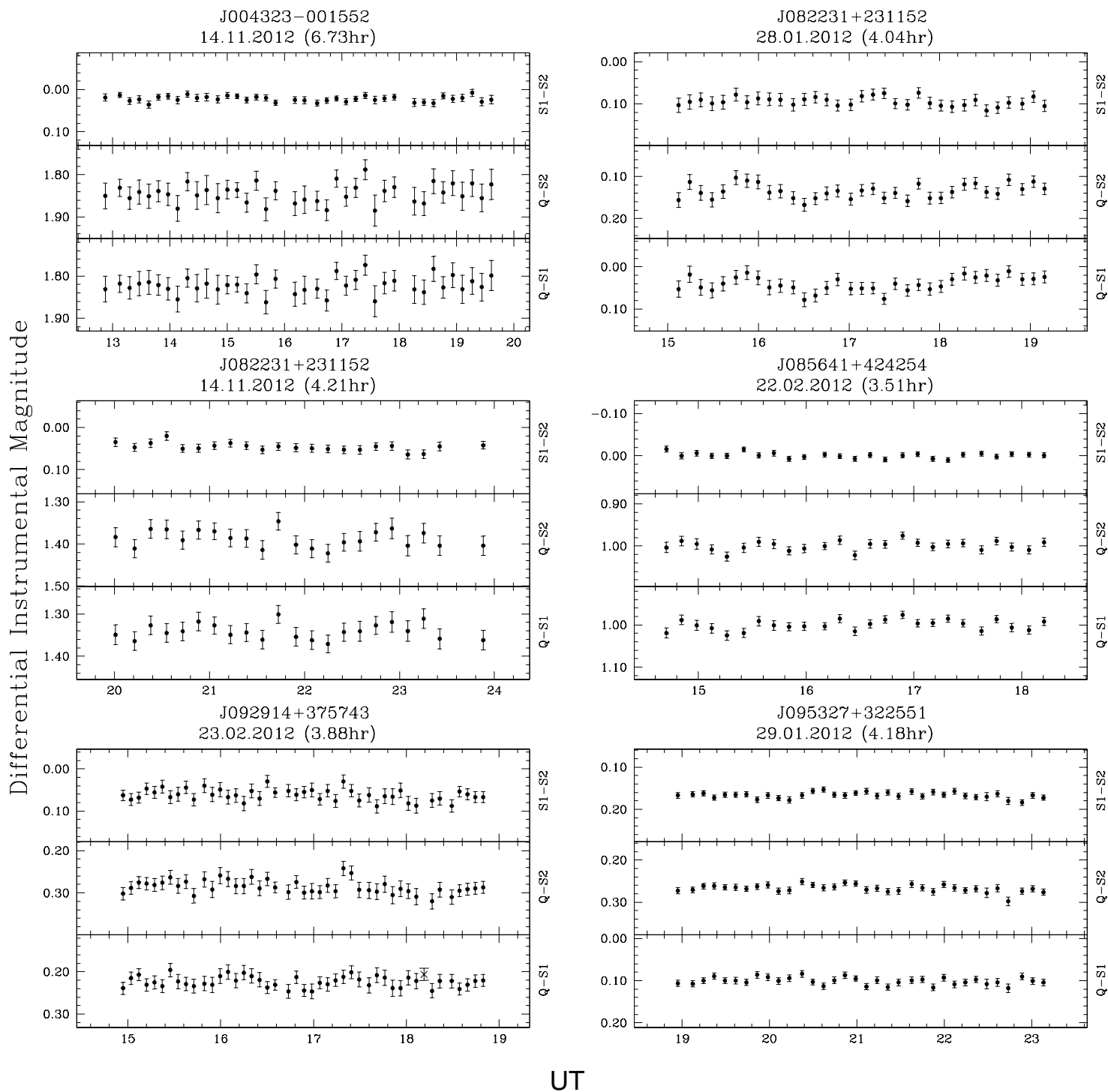,angle=00,bbllx=19bp,bblly=155bp,bburx=585bp,bbury=716bp,
clip=true} 
\caption{Differential light curves (DLCs) for the radio-loud BALQSOs
  in our sample. The name of the quasar and the date and duration of
  the observation are given at the top of each night's data. The upper
  panel gives the comparison star-star DLC and the subsequent lower
  panels give the quasar-star DLCs  as defined in the labels on the
  right side. Any likely outliers (at $> 3\sigma$) in the star-star
  DLCs are marked with crosses, and the data corresponding to such
  flagged exposure are also removed from quasar-star DLC, for the
  final analysis.}
\label{fig:balqso_dlurve} 
\end{figure*}

\begin{figure*} 
\centering
\epsfig{figure=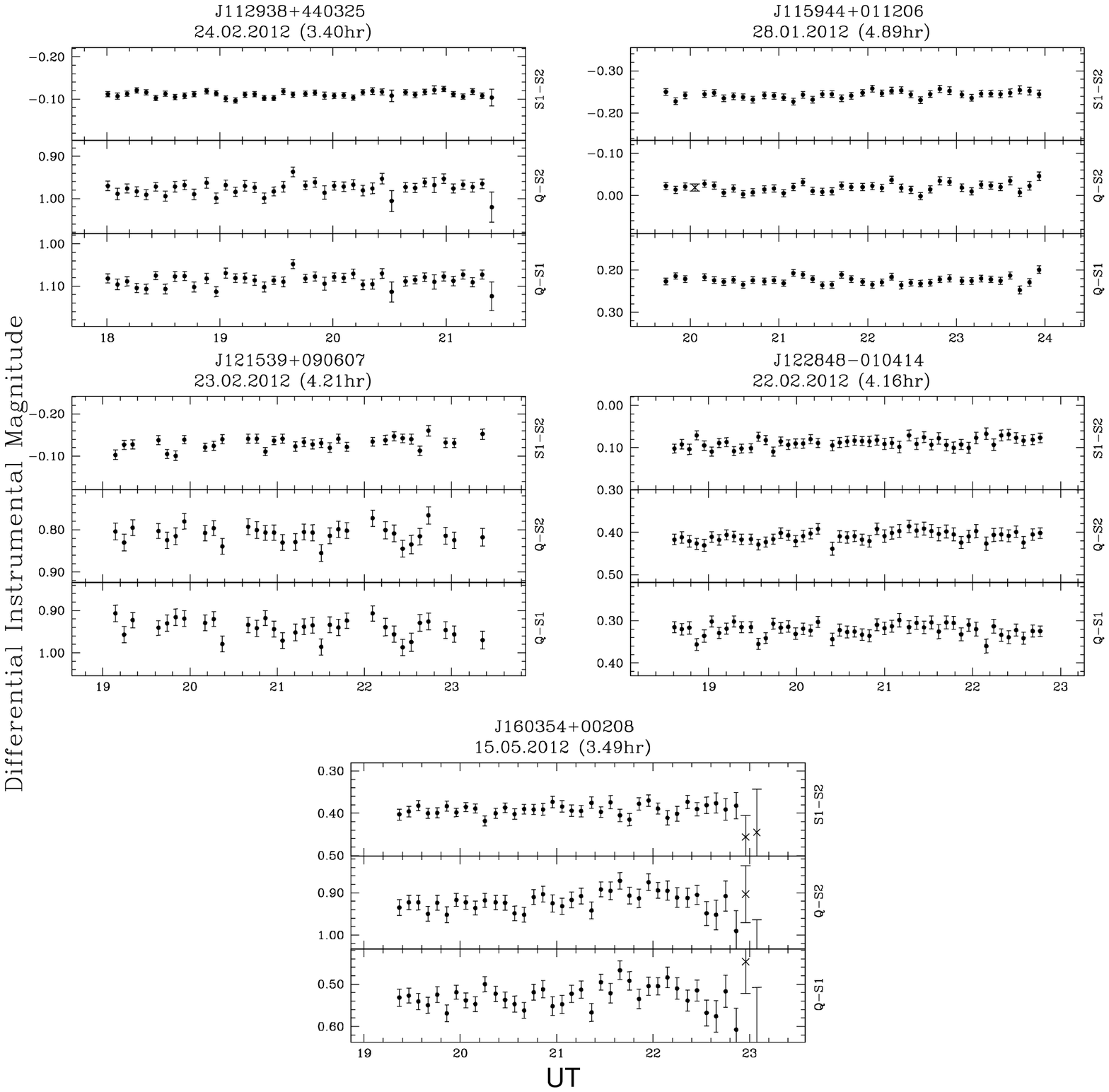,angle=00,bbllx=19bp,bblly=155bp,bburx=589bp,bbury=713bp,
clip=true} 
\caption{Same as in Figure 1.}
\label{fig:balqso_dlurve2} 
\end{figure*}

We employed a mean clip algorithm on the comparison star-star DLCs, so
as to remove any outliers, i.e. the sharp rise or fall of the DLC over
a single time bin which may arise due to improper removal of cosmic
rays or some unknown instrumental cause, above the 3$\sigma$ from the
mean. The data points corresponding to exposure resulting in such
outliers are also removed from the respective quasar-star DLCs as
well. However, we should stress here that such outliers in our comparison
star-star DLCs were usually not present and never exceeded two data
points. Finally, the statistical analysis of microvariability is
performed on the quasar-star and star-star DLCs, freed from such
outliers, as shown in Figure ~\ref{fig:balqso_dlurve}, \ref{fig:balqso_dlurve2}.

%%%%%%%%%%%%%%%%%%%%%%%%%%%%%%%%%%%%%
%%%%%%  Analysis \& Results  %%%%%%%%
%%%%%%%%%%%%%%%%%%%%%%%%%%%%%%%%%%%%%

\section{Analysis \& Results}
\label{sec:ananres_rlbal}

So far in the literature, the most commonly used statistics to
quantify the intra-night optical variability (INOV) of DLCs, is the
``\emph{C-statistics}'' ~\citep*{Romero1999A&AS..135..477R}, which is
the ratio of observational scatter of quasar-star and star-star DLCs.
However, ~\citet{Diego2010AJ....139.1269D} has pointed out that the C
is not a proper statistics, as the nominal confidence limit (i.e.
C$>$2.57) for the presence of variability with 99\% confidence, is too
conservative. \par

 ~\citet{Diego2010AJ....139.1269D} has shown that the other
alternatives to the $C$-statistics are the one-way analysis of
variance (ANOVA), $\chi^2 -$test and \emph{F$-$}test, which are much
powerful statistical tests to quantify the presence of
microvariability. We noticed that for an appropriate use of ANOVA, the
number of data points in the DLC needs to be large enough so as to
have many points in each subgroup used for the analysis; however, this
is not possible for our observations as we have typically only around
30-40 data points in our light curves. In addition, for the
appropriate use of a $\chi^{2} -$ test, the errors of individual data
points need to have Gaussian distribution and those errors should be
accurately estimated. It has been claimed in the literature that
errors returned by photometric reduction routines in IRAF and DAOPHOT
are usually underestimated, often by factors of 1.3$-$1.75
~\citep[][]{GopalKrishna2003ApJ...586L..25G, Sagar2004MNRAS.348..176S,
  Bachev2005MNRAS.358..774B, Goyal2012A&A...544A..37G}, which make
use of a $\chi^2 -$ test less desirable for such real photometric
light curves. Therefore, to quantify the INOV in DLCs, we prefer the
proper statistics that is reasonable to employ for differential
photometry, the \emph{F$-$}test ~\citep{Diego2010AJ....139.1269D}.
\par

Although the F-test is certainly better than the C-test, it should be
noted that for the F-test to give a truly reliable result, the error
due to random noise in the quasar-star and star-star DLCs should be of
a similar order, apart from any additional scatter in the quasar-star
DLC due to possible QSO variability. ~\citet{Joshi2011MNRAS.412.2717J}
have pointed out that the standard \emph{F$-$}test statistic does not
account for the brightness difference in between quasar and comparison
star as well as for the photometric error returned by the image
reduction routines, which can result the false alarm
detection(non-detection) of variability. For instance, if both
comparison stars are brighter(fainter) than the monitored quasar,
then a false alarm detection (non-detection) is possible due to the
very small(large) photon noise variance of the star-star DLC compared
to the quasar-star DLCs as demonstrated in
~\citet{Joshi2011MNRAS.412.2717J}. Although we have tried to select
our non-variable comparison star in proximity to both position and
magnitude to the quasar, but it is very difficult to find the same for
all the quasars. So here we have used the ``scaled \emph{F$-$}test''
statistic, as is suggested by ~\citet{Joshi2011MNRAS.412.2717J}, to
compute our \emph{F$-$}value as follows:

\begin{table*} 
\centering 
\caption{Positions and magnitudes of the radio-loud BALQSOs and the comparison stars. \label{tab_cdq_comp}}
%\tiny
\begin{tabular}{ccc ccc c}\\
\hline 

{IAU Name} &   Date       &   {R.A.(J2000)} & {Dec.(J2000)}                      & {\it g} & {\it r} & {\it g-r} \\  
           &  dd.mm.yy    &   (h m s)       &($^\circ$ $^\prime$ $^{\prime\prime}$)   & (mag)   & (mag)   & (mag)     \\  
{(1)}      & {(2)}        & {(3)}           & {(4)}                              & {(5)}   & {(6)}   & {(7)}     \\
\hline   
\multicolumn{7}{l}{}\\

J004323$-$001552 &  14.11.2012 & 00 43 23.43 & $-$00 15 52.4 & 18.76 & 18.47 &   0.29 \\
S1               &             & 00 43 45.79 & $-$00 19 31.6 & 16.99 & 16.55 &   0.44 \\
S2               &             & 00 433 9.19 & $-$00 20 24.6 & 16.88 & 16.49 &   0.39 \\    
J082231$+$231152 &  28.01.2012 & 08 22 31.53 & $+$23 11 52.0 & 17.93 &  17.68 &   0.25 \\
S1               &             & 08 23 09.66 & $+$23 14 42.4 & 17.37 &  16.85 &   0.52 \\
S2               &             & 08 22 12.58 & $+$23 05 07.4 & 18.15 &  16.85 &   1.30 \\
J082231$+$231152 &  14.11.2012 & 08 22 31.53 & $+$23 11 52.0 & 17.93 &  17.68 &   0.25 \\
S1               &             & 08 23 01.30 & $+$23 18 05.1 & 18.54 &  18.12 &   0.42 \\
S2               &             & 08 22 18.56 & $+$23 16 50.9 & 19.49 &  18.06 &   1.44 \\
J085641$+$424254 &  22.02.2012 & 08 56 41.56 & $+$42 42 53.9 & 18.81 &  18.50 &   0.31 \\
S1               &             & 08 56 31.48 & $+$42 37 51.2 & 17.77 &  17.49 &   0.28 \\
S2               &             & 08 56 19.09 & $+$42 35 09.5 & 18.12 &  17.37 &   0.75 \\
J092914$+$375743 &  23.02.2012 & 09 29 13.96 & $+$37 57 42.9 & 18.28 &  18.05 &   0.23 \\
S1               &             & 09 28 44.88 & $+$37 50 25.0 & 18.64 &  17.79 &   0.85 \\
S2               &             & 09 28 44.88 & $+$37 50 25.0 & 18.64 &  17.79 &   0.85 \\
J095327$+$322551 &  29.01.2012 & 09 53 27.95 & $+$32 25 51.6 & 17.77 &  17.36 &   0.41 \\
S1               &             & 09 52 59.50 & $+$32 29 47.0 & 18.21 &  17.12 &   1.09 \\
S2               &             & 09 52 55.29 & $+$32 33 30.3 & 17.91 &  17.30 &   0.61 \\
J112938$+$440325 &  24.02.2012 & 11 29 38.46 & $+$44 03 25.0 & 18.22 &  18.12 &   0.10 \\
S1               &             & 11 30 15.26 & $+$43 54 52.1 & 17.80 &  17.15 &   0.65 \\
S2               &             & 11 30 02.97 & $+$44 07 12.9 & 18.15 &  17.02 &   1.13 \\
J115944$+$011206 &  28.01.2012 & 11 59 44.82 & $+$01 12 06.9 & 17.58 &  17.25 &   0.33 \\
S1               &             & 11 59 48.28 & $+$01 19 38.6 & 17.55 &  17.16 &   0.39 \\
S2               &             & 12 00 07.71 & $+$01 10 33.0 & 17.93 &  17.38 &   0.56 \\
J121539$+$090607 &  23.02.2012 & 12 15 39.66 & $+$09 06 07.4 & 18.39 &  18.26 &   0.13 \\
S1               &             & 12 16 05.93 & $+$08 59 03.2 & 17.94 &  17.56 &   0.38 \\
S2               &             & 12 15 38.76 & $+$09 03 19.9 & 18.39 &  17.64 &   0.76 \\
J122848$-$010414 &  22.02.2012 & 12 28 48.21 & $-$01 04 14.5 & 18.28 &  18.17 &   0.11 \\
S1               &             & 12 28 21.82 & $-$01 02 36.4 & 19.31 &  17.85 &   1.46 \\
S2               &             & 12 28 22.35 & $-$01 01 19.2 & 17.86 &  17.65 &   0.22 \\
J160354$+$300208 &  15.05.2012 & 16 03 54.15 & $+$30 02 08.6 & 18.13 &  17.96 &   0.17 \\
S1               &             & 16 04 30.55 & $+$30 03 03.7 & 18.80 &  17.66 &   1.14 \\
S2               &             & 16 03 44.29 & $+$30 02 11.2 & 18.70 &  17.25 &   1.45 \\

\hline
\end{tabular}
\end{table*}

 \begin{table*}
  \centering
  \begin{minipage}{500mm}
 {
 \caption{INOV results for the radio-loud BALQSOs. }
 \label{tab:res}
 \begin{tabular}{@{}ccc cc ccr cc@{}} 
 \hline  \multicolumn{1}{c}{RL-BALQSO} &{Date} &{T} &{N} 
 &\multicolumn{3}{c}{\emph{scaled F$-$test}}
 &\multicolumn{1}{c}{Variability?{\footnote{V=variable, i.e. confidence
       $\ge 0.99$; PV=probable variable, i.e. $0.95-0.99$ confidence; NV
       =non-variable, i.e. confidence $< 0.95$.\\
 Variability status values based on quasar-star1 and quasar-star2
 pairs are separated by a comma.}}}
  &{$\sqrt\kappa${\footnote{Here
        $\kappa=\langle\sigma^2(q-s)\rangle/\langle\sigma^2(s1-s2)\rangle$ (as in Eq.~\ref{eq:kappa}), is
  used to scale the variance of star-star DLCs for  the \emph{scaled F$-$test}.}}}
 & {$\sqrt { \langle \sigma^2_{i,err} \rangle}$}\\  & dd.mm.yy& 
hrs & & {$F_1^{s}$},{$F_2^{s}$} & {$F_c{(0.95)}$} &{$F_c{(0.99)}$} & & &  \\ (1)&(2) &(3) &(4) &(5) &(6)
 &(7)&(8)&(9) &(10)\\ \hline

 J004323.42$-$001552.4  & 14.11.2012   & 6.73  & 38  &  0.76,0.94 & 1.73&2.18   & NV(NV, NV )   &   3.56   &   0.03\\  
 J082231.53$+$231152.0  & 28.01.2012   & 4.04  & 33  &  2.35,2.44 & 1.80&2.32   & V(V ,V   )    &   1.06   &   0.02 \\ 
 J082231.53$+$231152.0  & 14.11.2012   & 4.21  & 23  &  0.76,0.93 & 2.05&2.78   & NV(NV, NV )   &   2.31   &   0.02 \\ \\
 J085641.56$+$424253.9  & 22.02.2012   & 3.51  & 25  &  1.40,1.13 & 1.98&2.66   & NV(NV, NV )   &   1.67   &   0.01 \\ 
 J092913.96$+$375742.9  & 23.02.2012   & 3.88  & 44  &  0.78,1.08 & 1.66&2.06   & NV(NV, NV )   &   1.08   &   0.02 \\ 
 J095327.95$+$322551.6  & 29.01.2012   & 4.18  & 34  &  1.13,1.26 & 1.79&2.29   & NV(NV, NV )   &   1.14   &   0.01 \\ \\
 J112938.46$+$440325.0  & 24.02.2012   & 3.40  & 40  &  1.89,1.86 & 1.70&2.14   & PV(PV, PV )   &   1.75   &   0.01 \\ 
 J115944.82$+$011206.9  & 28.01.2012   & 4.89  & 44  &  2.03,1.92 & 1.66&2.06   & PV(PV, PV )   &   1.04   &   0.01 \\ 
 J121539.66$+$090607.4  & 23.02.2012   & 4.21  & 32  &  0.82,0.62 & 1.82&2.35   & NV(NV, NV )   &   1.77   &   0.02 \\ \\
 J122848.21$-$010414.4  & 22.02.2012   & 4.16  & 48  &  1.28,0.86 & 1.62&1.99   & NV(NV, NV )   &   1.18   &   0.01\\  
 J160354.14$+$300208.6  & 15.05.2012   & 3.49  & 36  &  2.35,1.91 & 1.76&2.23   & PV(V ,PV  )   &   1.49   &   0.02 \\ 
                                                                                                                  
\hline                                                                                                            
\end{tabular}                                                                                                     
}                                                                                                                 
\end{minipage}
\end{table*}

\begin{equation}
 F_{1}^{s}=\frac{Var(q-s1)}{\kappa \times Var(s1-s2)},  \\
 F_{2}^{s}=\frac{Var(q-s2)}{\kappa \times Var(s1-s2)}
\label{eq.fstest}
\end{equation}

with  $\kappa$, defined as,

\begin{equation}
\kappa=\left[\displaystyle{\frac{\sum_\mathbf{i=0}^\mathbf{N}\sigma^2_{i,err}(q-s)/N}
{\sum_\mathbf{i=0}^\mathbf{N}\sigma^2_{i,err}(s1-s2)/N}}\right] \equiv
\frac{\langle\sigma^2(q-s)\rangle}{\langle\sigma^2(s1-s2)\rangle}, \label{eq:kappa}
\end{equation} 

where $\sigma^2_{i,err}(q-s)$ and $\sigma^2_{i,err}(s1-s2)$ are, respectively, the errors
on individual data points of the quasar-star and star-star DLCs as returned by the
DAOPHOT routine. Here, the term $\kappa$ is a scaling factor, applied to the variance of
the star-star DLC. Firstly, this scaling factor will take care of the difference in
magnitude between the QSO and star in quasar-star and star-star DLCs. Secondly, it is free
from the problem of uncertain error underestimation by DAOPHOT/IRAF routines reported by
many other authors, because our scaling factor depends on the ratio of the averaged squared
errors and hence any factor with which errors are either
underestimated or overestimated will be canceled out. Therefore, we report
our final results based on this \emph{scaled F$-$test} (defined by $F^s$).\par 

We also note that, in standard F-test, the F-value is a ratio of the two sample
variance, with the assumption that error-bar in both the samples is of similar order. In other
words, more specifically when divided by $\sigma^2$ the variance will be distributed according
to $\chi^2$ distribution, and hence F-distribution by definition consists of the ratio of
two $\chi^2$ distributions, which is finally used to derive its probability distribution.
Our \emph{scaled F$-$test}, with $\kappa$ term, also  amount to the ratio of two $\chi^2$
distribution by assigning $\sigma^2=\sum \sigma^2_{i,err}/(N-1)$ as averaged error.
The only difference here, compared to the  standard  F-test, is 
that the error-bar appearing in  the $\chi^2$ numerator (i.e. of quasar-star DLC)
and denominator (i.e. of star-star DLC) do differ due to any brightness difference between  quasar and its  comparison stars. Therefore, in \emph{scaled F$-$test} 
unlike the standard F-test, $\sigma^2$ term  of the numerator and denominator 
though does not exactly get canceled out, but still do preserve its 
genuine F-distribution. \par

$F$ values computed by Eq ~\ref{eq.fstest}, are 
compared individually with the critical $F$ value,
$F^{(\alpha)}_{\nu_{QS}\nu_{SS}}$, where $\alpha$ is the significance
level set for the test, and $\nu_{QS}$ and $\nu_{SS}$ are the degrees
of freedom of the quasar-star and star-star DLCs, respectively. The
smaller the $\alpha$ value, the more improbable that the result be 
produced by chance. For the present study, we have used the
significance levels, $\alpha=$ 0.01 and 0.05, which corresponds to
confidence levels of greater than 99 and 95 per cent respectively. If
$F$ is larger than the critical value, the null hypothesis (i.e. no
variability) is discarded. Thus, for variability status we prefer to
compare the \emph{F$-$}value corresponding to quasar-star1 and
quasar-star2 DLCs (i.e. $F_1^s$ \& $F_2^s$) from Eq.~\ref{eq.fstest},
separately with the critical $F$ value, as if one DLC indicates
variability and other doesn't, these mixed signals bring into question
the reality of the putative variability. So, a quasar is marked as
\emph{Variable} (`V') for a $F$-value $\ge F_{c}(0.99)$, which
corresponds to a confidence level $\ge 0.99$; \emph{Probably variable}
(`PV') if the $F$-value is between $F_{c}(0.95)$ and
$F_{c}(0.99)$; \emph{Non variable} (`NV') if the $F$-value is less
than $F_{c}(0.95)$. Finally, we assign a source as; of class variable
(i.e. `V') if both the quasar-star1 and quasar-star2 states it as `V';
of class probably variable (i.e `PV') if either both DLCs show it as
`PV', or one of it as `PV' and other as `V'; of class non variable
(i.e. `NV') if any one of the quasar-star1 or quasar-star2 DLC show
it as non variable (i.e. `NV').

The DLCs of our radio-loud BALQSOs sample are shown in Figure
~\ref{fig:balqso_dlurve},~\ref{fig:balqso_dlurve2} and the corresponding results of our analysis
are summarized in Table~\ref{tab:res}. In the first four columns, we
list the object name, date of observation, the duration of our
observation, and the number of data points ($N_{points}$) in the DLC.
The column five gives the \emph{scaled F$-$test} values corresponding
to both quasar-star1 and quasar-star2 DLCs. The columns six, seven
mention the critical F-value for the corresponding 95 per cent and 99 per cent
significance level. In column eight, we give the variability status of
our source, based on the \emph{scaled F$-$test} for both the
quasar-star1 and quasar-star2 DLCs. Column nine lists the square root
of scaling factor, $\sqrt{\kappa}$, where $\kappa= \langle
\sigma^2(q-s)$/$\sigma^2(s1-s2) \rangle$ (as in
equation~\ref{eq.fstest}), and has been used to scale the variance of
the star$-$star DLCs while computing the F value in the \emph{scaled F$-$test}.
The last column gives our photometric accuracy,
$\sqrt { \langle \sigma^2_{i,err} \rangle}$  in the quasar$-$star DLCs, which is typically
of 0.01-0.03 mag.

%%%%%%%%%%%%%%%%%%%%%%%%%%%%%%%%%%%%%
%%%%%% The INOV duty cycle   %%%%%%%%
%%%%%%%%%%%%%%%%%%%%%%%%%%%%%%%%%%%%%

\subsection{The INOV duty cycle (\emph{DC})}
\label{label:dc}

The duty cycle of INOV was computed as is given by
~\citet*{Romero1999A&AS..135..477R},
\begin{equation} 
DC = 100\frac{\sum_\mathbf{i=1}^\mathbf{n} N_i(1/\Delta t_i)}{\sum_\mathbf{i=1}^\mathbf{n}(1/\Delta t_i)} {\rm per cent} 
\label{eq:dc} 
\end{equation} 

where $\Delta t_i = \Delta t_{i,obs}(1+z)^{-1}$ is the duration of the
monitoring session of a source on the $i^{th}$ night, corrected for
its cosmological redshift, $z$. Since the observing run time for a
given source is not same for different nights observation, so the
computation of DC has been weighted by the actual monitoring duration
$\Delta t_i$ on the $i^{th}$ night. Here, $N_i$ was set equal to 1  if
INOV was detected, otherwise $N_i$ = 0. \par

In 8 nights monitoring (i.e. for $\sim$46 hour) of our 10 RL-BALQSOs,
we find a clear signature of variability for one RL-BALQSO resulting
in INOV duty cycle (\emph{DC}) of 5.1 per cent. While on including the
three sources assigned as probable variable (`PV', class), the
\emph{DC} increases to 35.1 per cent. \par

Here, it is also worth comparing the DC of RL-BALQSOs with our earlier
study of 19 RQ-BALQSOs for microvariability properties
~\citep{Joshi2011MNRAS.412.2717J}. For that, we have computed the
INOV DC value of RQ-BALQSOs sample, by using Eq.~\ref{eq:dc}. An INOV DC
of 6.41 and 34.28 per cent was found respectively for clear detection
of variability (two out of 19 DLCs) and probably variable cases 
(six out of 19 DLCs), which is almost similar to our above DC of
RL-BALQSOs.

%%%%%%%%%%%%%%%%%%%%%%%%%%%%%%%%%%%%%
%%%%%% Discussion \& conclusions %%%%
%%%%%%%%%%%%%%%%%%%%%%%%%%%%%%%%%%%%%

\section{Discussion \& conclusions}
\label{sec:diss_con_rlbal}

Soon after the discovery of quasars, their behavior of changing flux in minutes to hours and
amplitude from few tenth to a few magnitude i.e. the microvariability, was
reported ~\citep{Matthews1963ApJ...138...30M}. Thereafter, plenty of efforts have been
made to correlate it with several other properties, such as quasar orientation, radio
loudness and polarization, driving the microvariation in various sub-classes of
AGN. The emerging picture is that the microvariability does not occur as often in
radio-quiet objects as it does in radio-loud sources
~\citep{GopalKrishna2003ApJ...586L..25G, Stalin2004MNRAS.350..175S,
Carini2007AJ....133..303C}. However, a multiband monitoring study of 22 radio-quiet and
22 core dominated radio-loud (CRL) quasars by ~\citet{Ramirez2009AJ....138..991R}, has
confirmed that the frequency of microvariability in both these quasar sub-classes are same;
similar to the earlier results found by ~\citet{Diego1998ApJ...501...69D}. Recently,
~\citet{Goyal2012A&A...544A..37G} has investigated the role of optical polarization along
with the relativistic beaming in microvariability property for a large sample of 21 core
dominated quasars (CDQs), consisting of 12 low polarization (LP) and 9 high polarization
(HP) QSOs. With the INOV duty cycle of 28 per cent and 68 per cent respectively for LPCDQs
and HPCDQs, they concluded that the relativistic beaming is normally not a sufficient
condition,  but high optical polarization is also a necessary condition for strong INOV.
\par

In general, the microvariability seen in radio-loud objects is now
widely believed to come from turbulence in a relativistic jet pointed
at or nearly along our line of sight. However, in the case of
radio-quiet objects, the jets are weaker than radio-loud jets or
quenched near their origin point due to the effects of black hole spin
~\citep{Wilson1995ApJ...438...62W,Blandford2000RSPTA.358..811B} or
magnetic configurations ~\citep{Meier2002NewAR..46..247M}. To know
which mechanism among these various possibilities is at play for
the origin of microvariability on different sub-classes of AGN, the
constraints obtained by extending the microvariability studies to
other remaining AGN sub-class such as BALQSOs have given important
clues ~\citep[see e.g.][]{Joshi2011MNRAS.412.2717J}. For instance,
the early evidence from spectro-polarimetry of BALQSOs give an idea that
they are viewed closer to the disk plane i.e. nearly edge-on
\citep*[e.g.][]{Goodrich1995ApJ...448L..73G}. It also supports the
models where outflow comes closer to the disk plane
\citep[e.g.][]{Elvis2000ApJ...545...63E}. It can be a clue to
understand the radio-loud/radio-quiet dichotomy as the BAL quasars
appeared to be exclusively radio-quiet, where outflows are not well
collimated. Later \citet{Brotherton2006MNRAS.372L..58B} have shown
that BALQSOs are viewed at a large variety of viewing angles 
($\sim 15{^\circ}$ from the jet axis to nearly edge-on). Constrained by
radio flux density variability to lie within $35{^\circ}$ of a
relativistic jet, give rise to a new population of polar BALQSO i.e. those
 viewed perpendicular to the disk
\citep*[e.g.][]{Zhou2006ApJ...639..716Z, Ghosh2007ApJ...661L.139G,
  Montenegro2008MNRAS.388.1853M, Doi2009PASJ...61.1389D}, where
outflows are confined to well-collimated relativistic jets. \par

In these contexts, here we have presented our results of intranight
optical variability observations of 10 extremely radio-loud (i.e. R$
>100$) BALQSOs, observed mostly in the first half of year 2012. These are the
first extensive observations of radio-loud sub-class of
BALQSOs with an aim to probe their microvariability nature. In other
words, we have investigated the effect of presence of BAL as well as
radio-loudness on microvariability properties. The application of
proper statistics, i.e. the ``\emph{scaled F$-$test}''
~\citep{Joshi2011MNRAS.412.2717J} has allowed us to find the genuine
confirmed INOV DC of 5.1 per cent (one variable DLC out of total 10),
which becomes 35.1 per cent if we also include the three probable variable (i.e.
`PV') cases as well. This DC is free from any spurious cause, as could
be  present at times while  using the standard F-test i.e. introduced by
the brightness difference in quasar and its comparison star (e.g. Section ~\ref{sec:ananres_rlbal}), 
since we have used \emph{scaled F$-$test}, 
described in detail in ~\citet{Joshi2011MNRAS.412.2717J}. Having used
the same statistical and analysis method, both in this work and
~\citet{Joshi2011MNRAS.412.2717J}, where they studied microvariation
properties of 19 radio-quiet BALQSOs, we carried out the comparison of
these two results and found that our resulting INOV DC of 5.1 per cent
matches with their result of 6.41 per cent fraction (where they found
out of 19 only 2 DLCs as variable, also see Section ~\ref{label:dc}).
This suggest that radio-loudness has not any significant role at least
in the case of microvariability properties of BALQSOs, but larger
sample size of radio-loud BALQSOs will help to say about it more
firmly.\par

As a general implication, if we would have also got high DC of
microvariation in our RL-BALQSOs, such as found by
\citet{Zhou2006ApJ...639..716Z} \& ~\citet{Ghosh2007ApJ...661L.139G}
based on their radio monitoring, then it would have suggested that the
BAL outflow are aligned very close to the jet axis (about $<35$ deg)
which may be responsible for the higher DC due to Doppler boosting. On
the other hand, the common wind based model of BAL outflow demand
the outflow to be preferentially equatorial when the accretion disk is
almost edge on ~\citep[e.g.][]{Elvis2000ApJ...545...63E}, predicting
the low DC as is also suggested by our this investigation. Though only
with such microvariability studies, it may be difficult to 
distinguish clearly among many possibilities in a quantitative manner; but with
our this study we certainly could conclude that the microvariation of
both radio-loud and radio-quiet BALQSOs are very similar, and the
level of microvariation occurrence in both cases is quite low as is
usually found for radio-quiet QSOs, the theoretical implication of
which for AGN unification scheme needs further exploration.

%%%%%%%%%%%%%%%%%%%%%%%%%%%%%%%%%%%%%
%%%%%%   Acknowledgments       %%%%%%
%%%%%%%%%%%%%%%%%%%%%%%%%%%%%%%%%%%%%

\section*{Acknowledgments}

The help rendered by Dr. B. Kumar and Dr. A. Omar during the
initial phase of this observing program, along with scientific staff
of 1.3-m DFOT telescope, Nainital, is gratefully acknowledged. We also
thanks Dr. A. C. Gupta for useful discussion and the anonymous referee 
for useful comments in improving the manuscript.

%%%%%%%%%%%%%%%%%%%%%%%%%%%%%%%%%%%%%%%%%%%%%%
%%%%  Brief notes on individual sources %%%%%%
%%%%%%%%%%%%%%%%%%%%%%%%%%%%%%%%%%%%%%%%%%%%%%

\appendix
\section{Brief notes on individual sources}
\label{sec:abt_source_rlbal}

\subsection{J004323.42$-$001552.5}
J004323.42$-$001552.5 is a high redshift (i.e. z=2.797) BALQSO
with a balnicity index\footnote{ The Balnicity Index
  metric is defined as
\begin{equation}
BI=-\int_{25000}^{3000}\left[1-\frac{f(v)}{0.9}\right]Cdv.\label{eq:1}
\end{equation}
Here, the limits of the integral are in units of $km~s^{-1}$, $f(v)$
is the normalized flux as a function of velocity displacement from
line center. The constant $C = 0$ everywhere, unless the normalized
flux has satisfied $f(v) < 0.9$ continuously for at least $2000$ ${\rm
 km~s^{-1}}$, at which point it is switched to $C = 1$ until $f(v) >
0.9$ again. Based on this definition, objects are classified as
BALQSOs if their $BI>0$ ${\rm km~s^{-1}}$.}(BI) of 591.10 \kms
~\citep{Gibson2009ApJ...692..758G}. It appears as non-variable according to our \emph{scaled F$-$test} with an observation length of ~6.73 hr.

\subsection{J082231.53$+$231152.0}
J082231.53$+$231152.0 is a low redshift (i.e. z=0.653), \mgii low
ionization broad absorption line (loBAL) quasar
~\citep{Zhang2010ApJ...714..367Z}. This source has been monitored twice,
firstly on the night of 28 January 2012 for about 4.04 hr and secondly 
on the night of 14 November 2012 for about 4.21 hr. 
On applying the \emph{scaled F$-$test} statistics, this source has indicated a clear signature of
microvariability in its first DLC  lasting $\sim$4.04 hr while appeared 
as non-variable in its second DLC (e.g see Table~\ref{tab:res}).  This 
 makes it a promising candidate for future INOV monitoring.

\subsection{J085641.56$+$424253.9}

J085641.56$+$424253.9 is a \civ high ionization broad absorption line
(HiBAL) quasar, with a balnicity index of 820.60 \kms
~\citep{Trump2006ApJS..165....1T,Gibson2009ApJ...692..758G}. This
BALQSO is also identified as a high redshift, z=3.061, weak emission
line quasar by ~\citet{DiamondStanic2009ApJ...699..782D}. We observed
this source for $\sim$3.51 hr duration for which the statistical 
analysis of its DLC does not give any clue of rapid variability.

\subsection{J092913.96$+$375742.9}

This source has a massive black hole of mass 8.9$\times 10^{9} M_{\odot}$, 
estimated using its \mgii  line width ~\citep[see
  e.g.][]{Shen2008ApJ...680..169S}. This source is a HiBAL with BI
value corresponding to its \civ line is 1204.7 \kms. We found this
source to be non-variable during our ~3.9 hr observation.

\subsection{J095327.95$+$322551.6}

This source is a flat spectrum radio quasar type with a 4.8 GHz flux
density of 184mJy ~\citep{Healey2007ApJS..171...61H}. This source is
detected as BAL in ~\citet{ Gibson2009ApJ...692..758G} SDSS DR-5 BAL
quasar catalog, with a \civ balnicity index of 267.0 \kms. This source
has not shown any signature of microvariability, for an observing run
of 4.18 hr.

\subsection{J112938.46$+$440325.0}

This BAL quasar has a \civ BI value of 806.5 \kms
~\citep{Gibson2009ApJ...692..758G} and has a massive black hole of,
1.32$\times 10^{10} M_{\odot}$ as estimated by using the \mgii line
width ~\citep{Shen2008ApJ...680..169S}. This source was found to be
probably variable during the course of our ~ 3.40 hr observation,
which makes it a potential source for further microvariability study.

\subsection{J115944.82$+$011206.9}

This source, also known as, PKS 1157+014, has long been known to
possess a 21 cm absorption at $z_{abs} =
1.94$~\citep{Wolfe1986ApJS...61..249W} associated with the DLA
~\citep{Colbert2002ApJ...566...51C}.
~\citet{Stocke1984ApJ...280..476S} found it as unresolved point source
in 488MHz VLA observation, with flux value of 108mJy. It is a HiBAL
with balnicity index of 299.3, 632.9 \kms, respectively for \siiv and
\civ~\citep{Gibson2009ApJ...692..758G}. This source is a promising
candidate for future investigation as it appears probably variable
for our observing run of 4.89 hr.

\subsection{J121539.66$+$090607.4}

This source is a member of large bright quasar survey and is also known as LBQS
1213+0922. It has a massive black hole of mass $5.12\times 10^{9}
M_{\odot}$ estimated by using the \civ  line width by
~\citet{Shen2008ApJ...680..169S}. This is a narrow trough high
ionization BAL (i.e. nHiBAL) QSO with BI value 116.30 \kms, estimated by
\civ line. During the monitoring of this source, few exposures got corrupted (about maximum 8-9) due to some technical problem, which we have removed manually resulting in few gaps in its DLC.
Over the monitoring duration of $\sim$4.21 hr, we found this source to
be non-variable.

\subsection{J122848.21$-$010414.4}

This is a HiBAL quasar, with BI value of 17.0 \kms, corresponding to
the \civ line. It appears as non-variable according to our \emph{scaled F$-$test}
with an observation length of ~4.16 hr.

\subsection{J160354.14$+$300208.6}

This is a HiBAL quasar with a massive black hole of mass $1.46\times
 10^{9} M_{\odot}$ estimated by using the \mgii
 line~\citep{Shen2008ApJ...680..169S}. Its BI value is 84.7, 480.0
 \kms, respectively for \siiv and \civ line. Over an observation
 length of 3.49 hr, we found it as probably variable (i.e. `PV'), which
 makes it a potential candidate for further microvariability study.

\label{lastpage}

\bibliography{references}

\begin{thebibliography}{47}
\expandafter\ifx\csname natexlab\endcsname\relax\def\natexlab#1{#1}\fi

\bibitem[{Abazajian} et~al.(2009){Abazajian}, {Adelman-McCarthy}, {Ag{\"u}eros}
  et~al.]{Abazajian2009ApJS..182..543A}
{Abazajian} K.~N., {Adelman-McCarthy} J.~K., {Ag{\"u}eros} M.~A., et~al., 2009,
  \apjs, 182, 543

\bibitem[{Bachev} et~al.(2005){Bachev}, {Strigachev} \&
  {Semkov}]{Bachev2005MNRAS.358..774B}
{Bachev} R., {Strigachev} A., {Semkov} E., 2005, \mnras, 358, 774

\bibitem[{Becker} et~al.(2000){Becker}, {White}, {Gregg}, {Brotherton},
  {Laurent-Muehleisen} \& {Arav}]{Becker2000ApJ...538...72B}
{Becker} R.~H., {White} R.~L., {Gregg} M.~D., {Brotherton} M.~S.,
  {Laurent-Muehleisen} S.~A., {Arav} N., 2000, \apj, 538, 72

\bibitem[{Blandford}(2000)]{Blandford2000RSPTA.358..811B}
{Blandford} R.~D., 2000, in { Astronomy, physics and chemistry of
  H$^{+}$$_{3}$\/}, vol. 358 of { Royal Society of London Philosophical
  Transactions Series A\/},  811--829

\bibitem[{Brotherton} et~al.(2006){Brotherton}, {De Breuck} \&
  {Schaefer}]{Brotherton2006MNRAS.372L..58B}
{Brotherton} M.~S., {De Breuck} C., {Schaefer} J.~J., 2006, \mnras, 372, L58

\bibitem[{Carini} et~al.(1992){Carini}, {Miller}, {Noble} \&
  {Goodrich}]{Carini1992AJ....104...15C}
{Carini} M.~T., {Miller} H.~R., {Noble} J.~C., {Goodrich} B.~D., 1992, \aj,
  104, 15

\bibitem[{Carini} et~al.(2007){Carini}, {Noble}, {Taylor} \&
  {Culler}]{Carini2007AJ....133..303C}
{Carini} M.~T., {Noble} J.~C., {Taylor} R., {Culler} R., 2007, \aj, 133, 303

\bibitem[{Colbert} \& {Malkan}(2002)]{Colbert2002ApJ...566...51C}
{Colbert} J.~W., {Malkan} M.~A., 2002, \apj, 566, 51

\bibitem[{de Diego}(2010)]{Diego2010AJ....139.1269D}
{de Diego} J.~A., 2010, \aj, 139, 1269

\bibitem[{de Diego} et~al.(1998){de Diego}, {Dultzin-Hacyan}, {Ramirez} \&
  {Benitez}]{Diego1998ApJ...501...69D}
{de Diego} J.~A., {Dultzin-Hacyan} D., {Ramirez} A., {Benitez} E., 1998, \apj,
  501, 69

\bibitem[{Diamond-Stanic} et~al.(2009){Diamond-Stanic}, {Fan}, {Brandt}
  et~al.]{DiamondStanic2009ApJ...699..782D}
{Diamond-Stanic} A.~M., {Fan} X., {Brandt} W.~N., et~al., 2009, \apj, 699, 782

\bibitem[{Doi} et~al.(2009){Doi}, {Kawaguchi}, {Kono}
  et~al.]{Doi2009PASJ...61.1389D}
{Doi} A., {Kawaguchi} N., {Kono} Y., et~al., 2009, \pasj, 61, 1389

\bibitem[{Elvis}(2000)]{Elvis2000ApJ...545...63E}
{Elvis} M., 2000, \apj, 545, 63

\bibitem[{Ghosh} \& {Punsly}(2007)]{Ghosh2007ApJ...661L.139G}
{Ghosh} K.~K., {Punsly} B., 2007, \apjl, 661, L139

\bibitem[{Gibson} et~al.(2009){Gibson}, {Jiang}, {Brandt}
  et~al.]{Gibson2009ApJ...692..758G}
{Gibson} R.~R., {Jiang} L., {Brandt} W.~N., et~al., 2009, \apj, 692, 758

\bibitem[{Goodrich} \& {Miller}(1995)]{Goodrich1995ApJ...448L..73G}
{Goodrich} R.~W., {Miller} J.~S., 1995, \apjl, 448, L73

\bibitem[{Gopal-Krishna} et~al.(1993){Gopal-Krishna}, {Sagar} \&
  {Wiita}]{GopalKrishna1993MNRAS.262..963G}
{Gopal-Krishna}, {Sagar} R., {Wiita} P.~J., 1993, \mnras, 262, 963

\bibitem[{Gopal-Krishna} et~al.(2003){Gopal-Krishna}, {Stalin}, {Sagar} \&
  {Wiita}]{GopalKrishna2003ApJ...586L..25G}
{Gopal-Krishna}, {Stalin} C.~S., {Sagar} R., {Wiita} P.~J., 2003, \apjl, 586,
  L25

\bibitem[{Goyal} et~al.(2012){Goyal}, {Gopal-Krishna}, {Wiita}
  et~al.]{Goyal2012A&A...544A..37G}
{Goyal} A., {Gopal-Krishna}, {Wiita} P.~J., et~al., 2012, \aap, 544, A37

\bibitem[{Green} et~al.(2001){Green}, {Aldcroft}, {Mathur}, {Wilkes} \&
  {Elvis}]{Green2001ApJ...558..109G}
{Green} P.~J., {Aldcroft} T.~L., {Mathur} S., {Wilkes} B.~J., {Elvis} M., 2001,
  \apj, 558, 109

\bibitem[{Gupta} \& {Joshi}(2005)]{Gupta2005A&A...440..855G}
{Gupta} A.~C., {Joshi} U.~C., 2005, \aap, 440, 855

\bibitem[{Healey} et~al.(2007){Healey}, {Romani}, {Taylor}
  et~al.]{Healey2007ApJS..171...61H}
{Healey} S.~E., {Romani} R.~W., {Taylor} G.~B., et~al., 2007, \apjs, 171, 61

\bibitem[{Hewett} \& {Foltz}(2003)]{Hewett2003AJ....125.1784H}
{Hewett} P.~C., {Foltz} C.~B., 2003, \aj, 125, 1784

\bibitem[{Jiang} et~al.(2007){Jiang}, {Fan}, {Ivezi{\'c}}
  et~al.]{Jiang2007ApJ...656..680J}
{Jiang} L., {Fan} X., {Ivezi{\'c}} {\v Z}., et~al., 2007, \apj, 656, 680

\bibitem[{Joshi} et~al.(2011){Joshi}, {Chand}, {Gupta} \&
  {Wiita}]{Joshi2011MNRAS.412.2717J}
{Joshi} R., {Chand} H., {Gupta} A.~C., {Wiita} P.~J., 2011, \mnras, 412, 2717

\bibitem[{Matthews} \& {Sandage}(1963)]{Matthews1963ApJ...138...30M}
{Matthews} T.~A., {Sandage} A.~R., 1963, \apj, 138, 30

\bibitem[{Meier}(2002)]{Meier2002NewAR..46..247M}
{Meier} D.~L., 2002, \nar, 46, 247

\bibitem[{Miller} et~al.(1990){Miller}, {Peacock} \&
  {Mead}]{Miller1990MNRAS.244..207M}
{Miller} L., {Peacock} J.~A., {Mead} A.~R.~G., 1990, \mnras, 244, 207

\bibitem[{Montenegro-Montes} et~al.(2008){Montenegro-Montes}, {Mack}, {Vigotti}
  et~al.]{Montenegro2008MNRAS.388.1853M}
{Montenegro-Montes} F.~M., {Mack} K.-H., {Vigotti} M., et~al., 2008, \mnras,
  388, 1853

\bibitem[{Ram{\'{\i}}rez} et~al.(2009){Ram{\'{\i}}rez}, {de Diego}, {Dultzin}
  \& {Gonz{\'a}lez-P{\'e}rez}]{Ramirez2009AJ....138..991R}
{Ram{\'{\i}}rez} A., {de Diego} J.~A., {Dultzin} D., {Gonz{\'a}lez-P{\'e}rez}
  J.-N., 2009, \aj, 138, 991

\bibitem[{Reichard} et~al.(2003){Reichard}, {Richards}, {Hall}
  et~al.]{Reichard2003AJ....126.2594R}
{Reichard} T.~A., {Richards} G.~T., {Hall} P.~B., et~al., 2003, \aj, 126, 2594

\bibitem[{Romero} et~al.(1999){Romero}, {Cellone} \&
  {Combi}]{Romero1999A&AS..135..477R}
{Romero} G.~E., {Cellone} S.~A., {Combi} J.~A., 1999, \aaps, 135, 477

\bibitem[{Sagar} et~al.(2011){Sagar}, {Omar}, {Kumar}
  et~al.]{Sagar2011Csi...101...8.25}
{Sagar} R., {Omar} A., {Kumar} B., et~al., 2011, CURRENT SCIENCE, 101, 8

\bibitem[{Sagar} et~al.(2004){Sagar}, {Stalin}, {Gopal-Krishna} \&
  {Wiita}]{Sagar2004MNRAS.348..176S}
{Sagar} R., {Stalin} C.~S., {Gopal-Krishna}, {Wiita} P.~J., 2004, \mnras, 348,
  176

\bibitem[{Schneider} et~al.(2010{\natexlab{a}}){Schneider}, {Richards}, {Hall}
  et~al.]{Schneider2010yCat.7260....0S}
{Schneider} D.~P., {Richards} G.~T., {Hall} P.~B., et~al., 2010{\natexlab{a}},
  VizieR Online Data Catalog, 7260, 0

\bibitem[{Schneider} et~al.(2010{\natexlab{b}}){Schneider}, {Richards}, {Hall}
  et~al.]{Schneider2010AJ....139.2360S}
{Schneider} D.~P., {Richards} G.~T., {Hall} P.~B., et~al., 2010{\natexlab{b}},
  \aj, 139, 2360

\bibitem[{Shen} et~al.(2008){Shen}, {Greene}, {Strauss}, {Richards} \&
  {Schneider}]{Shen2008ApJ...680..169S}
{Shen} Y., {Greene} J.~E., {Strauss} M.~A., {Richards} G.~T., {Schneider}
  D.~P., 2008, \apj, 680, 169

\bibitem[{Shen} et~al.(2011){Shen}, {Richards}, {Strauss}
  et~al.]{Shen2011ApJS..194...45S}
{Shen} Y., {Richards} G.~T., {Strauss} M.~A., et~al., 2011, \apjs, 194, 45

\bibitem[{Stalin} et~al.(2004){Stalin}, {Gopal-Krishna}, {Sagar} \&
  {Wiita}]{Stalin2004MNRAS.350..175S}
{Stalin} C.~S., {Gopal-Krishna}, {Sagar} R., {Wiita} P.~J., 2004, \mnras, 350,
  175

\bibitem[{Stetson}(1987)]{Stetson1987PASP...99..191S}
{Stetson} P.~B., 1987, \pasp, 99, 191

\bibitem[{Stocke} et~al.(1984){Stocke}, {Foltz}, {Weymann} \&
  {Christiansen}]{Stocke1984ApJ...280..476S}
{Stocke} J.~T., {Foltz} C.~B., {Weymann} R.~J., {Christiansen} W.~A., 1984,
  \apj, 280, 476

\bibitem[{Stocke} et~al.(1992){Stocke}, {Morris}, {Weymann} \&
  {Foltz}]{Stocke1992ApJ...396..487S}
{Stocke} J.~T., {Morris} S.~L., {Weymann} R.~J., {Foltz} C.~B., 1992, \apj,
  396, 487

\bibitem[{Trump} et~al.(2006){Trump}, {Hall}, {Reichard}
  et~al.]{Trump2006ApJS..165....1T}
{Trump} J.~R., {Hall} P.~B., {Reichard} T.~A., et~al., 2006, \apjs, 165, 1

\bibitem[{Wilson} \& {Colbert}(1995)]{Wilson1995ApJ...438...62W}
{Wilson} A.~S., {Colbert} E.~J.~M., 1995, \apj, 438, 62

\bibitem[{Wolfe} et~al.(1986){Wolfe}, {Turnshek}, {Smith} \&
  {Cohen}]{Wolfe1986ApJS...61..249W}
{Wolfe} A.~M., {Turnshek} D.~A., {Smith} H.~E., {Cohen} R.~D., 1986, \apjs, 61,
  249

\bibitem[{Zhang} et~al.(2010){Zhang}, {Wang}, {Wang}, {Zhou}, {Dong} \&
  {Wang}]{Zhang2010ApJ...714..367Z}
{Zhang} S., {Wang} T.-G., {Wang} H., {Zhou} H., {Dong} X.-B., {Wang} J.-G.,
  2010, \apj, 714, 367

\bibitem[{Zhou} et~al.(2006){Zhou}, {Wang}, {Wang}, {Wang}, {Yuan} \&
  {Lu}]{Zhou2006ApJ...639..716Z}
{Zhou} H., {Wang} T., {Wang} H., {Wang} J., {Yuan} W., {Lu} Y., 2006, \apj,
  639, 716

\end{thebibliography}
\end{document}